\documentstyle[12pt,epsfig]{article}
\begin{document}
\rightline{IFUP-TH/2007-30}

\vskip 1cm

\centerline{{\bf Mass gap in the 2D O(3) non--linear sigma model with a $\theta=\pi$ term}}
\vskip 5mm
\centerline{B.~All\'es$^{\rm a}$, A. Papa$^{\rm b}$}

\centerline{\it $^{\rm a}$INFN Sezione di Pisa, Pisa, Italy}

\centerline{\it $^{\rm b}$Dipartimento di Fisica, Universit\`a della Calabria and}
\centerline{\it  INFN Gruppo Collegato di Cosenza, Arcavacata di Rende, Cosenza, Italy}

\begin{abstract}
By analytic continuation to real $\theta$ of data
obtained from numerical simulation at imaginary $\theta$
we study the Haldane conjecture and
show that the O(3) non--linear sigma model with
a $\theta$ term in 2 dimensions becomes massless at
$\theta=3.10(5)$. A modified cluster algorithm has been
introduced to simulate the model with imaginary $\theta$.
Two different definitions of the topological charge on the
lattice have been used; one of them needs renormalization
to match the continuum operator. Our work also offers a
successful test for numerical methods based on analytic
continuation.
\end{abstract}

\vfill\eject


\section{Introduction}

Several years ago Zamolodchikov and Zamolodchikov introduced
an integrable $S$--matrix for massless particles which was
associated to the 2--dimen-\break sional O(3) non--linear sigma model with
a topological $\theta$ term for $\theta=\pi$~\cite{zamolodchikov}.

Therefore the 2--dimensional O(3) sigma model with $\theta=\pi$
is possibly
gapless. Actually this conclusion was previously achieved by Haldane
and Affleck. They worked out a low energy description of the
1--dimensional chain of quantum half--integer spin with
antiferromagnetic coupling
finding that it and the above--mentioned O(3) sigma model
share the same long distance properties~\cite{haldane1,affleck1,shankar}.
Moreover it was found that antiferromagnetic quantum spin
chains are gapless for half--integer spins~\cite{haldane1,affleck2}.

Later Affleck and Haldane~\cite{haldane2} argued that the
critical theory for the half--integer quantum antiferromagnetic
spin chain is the Wess--Zumino--Witten model with a topological
coupling $k=1$. This model is the stable fixed point of the
2--dimensional O(3) sigma model with a vacuum angle $\theta=\pi$.

In addition, two numerical calculations of the partition
function for the O(3) model in the presence of a $\theta$
term~\cite{bietenholz} yield indications that the theory
undergoes a second order phase transition at $\theta=\pi$ (although
the two analyses disagree about the universality class).

In the present work we introduce a direct numerical method
to verify the Haldane conjecture for the 2--dimensional O(3) non--linear
sigma model. The idea is to perform a Monte Carlo study on the
lattice of the mass gap in the model as a function of
the $\theta$ parameter and to show that it vanishes
at a precise value of $\theta$, called $\theta_{\rm end}$,
which, following Haldane, should be $\theta_{\rm end}=\pi$.
We overcome the sign problem by simulating the theory at imaginary
$\theta$ and analytically continuing the results to the real
$\theta$ values. To this end we introduce a new cluster algorithm that
works for imaginary non--zero theta.

\section{Lattice implementation}

The continuum expression for the
action of the model is
\begin{eqnarray}
 S&=&A - i \theta Q\;, \quad A= \frac{1}{2g}
       \int \hbox{d}^2x \left(\partial_\mu \vec{\phi}(x)\right)^2\;,
       \nonumber \\
 Q&=&\int \hbox{d}^2x \,Q(x)\;,\quad \nonumber \\
 Q(x)&\equiv&
      \frac{1}{8\pi} \epsilon^{\mu\nu} \epsilon_{abc}
      \phi^a(x) \partial_\mu \phi^b(x)
      \partial_\nu \phi^c(x)\; ,
\label{continuumaction}
\end{eqnarray}
where $g$ is the coupling constant and $Q(x)$ is the topological charge
density. $\vec{\phi}(x)$ is a 3--component unit vector that represents
the dynamical variable at the site $x$. We have regularized this action
on a square lattice by the expression
\begin{equation}
 S_L = A_L - i \theta_L Q_L \;, \quad
 A_L \equiv-\beta \sum_{x,\mu} \vec{\phi}(x)
 \cdot \vec{\phi}(x+\widehat{\mu}) \;,
\label{latticeaction}
\end{equation}
where $Q_L=\sum_xQ_L(x)$ is the total lattice topological charge
and $Q_L(x)$ is the lattice topological charge density. $\beta$ is the
inverse bare lattice coupling constant and $\theta_L$
is the bare vacuum angle. In general, $\theta_L\not=\theta$ and
the point where the lattice regularized model becomes massless
will be called $\theta_{L,{\rm end}}$.

The action $A_L$ used in~(\ref{latticeaction}) is the simplest one on
the lattice. More complicated actions and expansion parameters boast
better scaling and asymptotic scaling properties and hence
they are more suited for the calculation of
masses~\cite{parisi,symanzik,alles}.
However our interest lies only on the vanishing of the mass gap
at a particular value of $\theta_L$ and such a property is clearly
unaffected by the slow convergence of the series.

\section{Choice of $Q_L$}

Let us discuss the regularization of the
topological charge density to be used in our Monte Carlo simulation.
$Q$ counts how many times the configuration of spin variables winds
around the unit sphere. Hence $Q$ takes on integer values.
Configurations with $+1$ ($-1$) winding number are called
instantons (anti--instantons)~\cite{belavin}.

We have made use of two different lattice regularizations
for the topological charge
density. The first one~\cite{papa}
\begin{eqnarray}
Q_L^{(1)}(x)&\equiv&\frac{1}{32\pi} \epsilon^{\mu\nu} \epsilon_{abc}\phi^a(x)
       \Big(\phi^b(x + \widehat{\mu}) - \phi^b(x - \widehat{\mu})\Big)
    \cdot \nonumber \\
     &&  \Big(\phi^c(x + \widehat{\nu}) - \phi^c(x - \widehat{\nu})\Big)\; ,
\label{latticeQ1}
\end{eqnarray}
is a symmetrical discretization of the expression 
for $Q(x)$ in Eq.(\ref{continuumaction}).

The second lattice regularization
is defined on triangles
(not on single sites). Every plaquette of a square lattice can be
cut through a diagonal into two triangles. If we call $\vec{\phi}_1$,
$\vec{\phi}_2$ and $\vec{\phi}_3$ the fields at the sites of the
three vertices (numbered counter--clockwise) of one of these triangles
then the fraction of spherical angle subtended by these fields is
$Q^{(2)}_L(\bigtriangleup)$ which satisfies~\cite{berg}
\begin{equation}
\exp\left(2\pi i  Q_L^{(2)}(\bigtriangleup)\right) =
     \frac{1}{\rho}\Big(
     1+\vec{\phi}_1 \cdot \vec{\phi}_2 +
        \vec{\phi}_2 \cdot \vec{\phi}_3 +
        \vec{\phi}_3 \cdot \vec{\phi}_1 +\nonumber
      i \vec{\phi}_1 \cdot \left(\vec{\phi}_2 \times
\vec{\phi_3}\right)\Big)\,,
\label{latticeQ2}
\end{equation}
where $\rho^2\equiv 2 (1 + \vec{\phi}_1 \cdot \vec{\phi}_2)
(1 + \vec{\phi}_2 \cdot \vec{\phi}_3)
(1 + \vec{\phi}_3 \cdot \vec{\phi}_1)$ and
$Q_L^{(2)}(\bigtriangleup)\in [-\frac{1}{2},+\frac{1}{2}]$.
The sum of $Q_L^{(2)}(\bigtriangleup)$ over all triangles yields
the so--called geometric topological charge~$Q_L^{(2)}$.

In general, a regularization of $Q$ does not lead to integer values
on a single configuration. To recover integer results for $Q_L$ on ensembles
of configurations that belong to the same topological sector,
we must renormalize this operator. The lattice and
the continuum topological charges are related by~\cite{pisa1}
\begin{equation}
 Q_L^{(1,2)}=Z_Q^{(1,2)}Q\,,
\label{qlzq}
\end{equation}
$Z_Q^{(1,2)}$ being the corresponding renormalization constant which
is UV finite since the topological charge operator has no anomalous
dimensions in the model under study.

$Z_Q^{(1,2)}$ can be calculated either in perturbation
theory~\cite{pisa1} or by a non--perturbative numerical method~\cite{pisa2}.
We have used the latter. In a nutshell it works in the following way:
a classical instanton (with topological charge +1) is put by hand on the
lattice and then 100 updating steps are applied (we used the Heat--Bath
algorithm on the conventional O(3) non--linear $\sigma$ model without a $\theta$
term since the renormalization constant to be used in Eq.(\ref{qlzq}) cannot
depend on $\theta$). After every Heat--Bath step the value of $Q_L^{(1,2)}$
is measured and, in order to monitor the background charge and check that
it is not varied after the updating step, $Q_L^{(1,2)}$ is measured again
after 6~cooling hits. In the calculation of $Z_Q^{(1)}$ this procedure was repeated
$4\cdot 10^4$ times at $\beta=1.5$ and $1.6$ and $10^4$ times for
$\beta=1.7$ and $1.75$. The average of $Q_L^{(1,2)}$ on configurations
within the topological sector $+1$ yields $Z_Q^{(1,2)}$.

The above non--perturbative method is summarized by the expression
\begin{equation}
Z_Q^{(1,2)}=\frac{\int_{\rm 1-instanton} {\cal D}{\vec{\phi}}\; Q_L^{(1,2)}\;
\exp\left(-A_L\right)}{\int_{\rm 1-instanton} {\cal D}{\vec{\phi}}\;
\exp\left(-A_L\right)}\,.
\label{heatingmethod}
\end{equation}
The restricted path integral runs over all
configurations (fluctuations) that preserve the background of one instanton.
Since the geometric charge $Q_L^{(2)}$ is +1 till the background classical
configuration is one instanton (whatever the fluctuations are), the
expression~(\ref{heatingmethod}) yields $Z_Q^{(2)}=1$ for all $\beta$~\cite{luscher}.

\begin{figure}[htbp]
\centerline{\epsfig{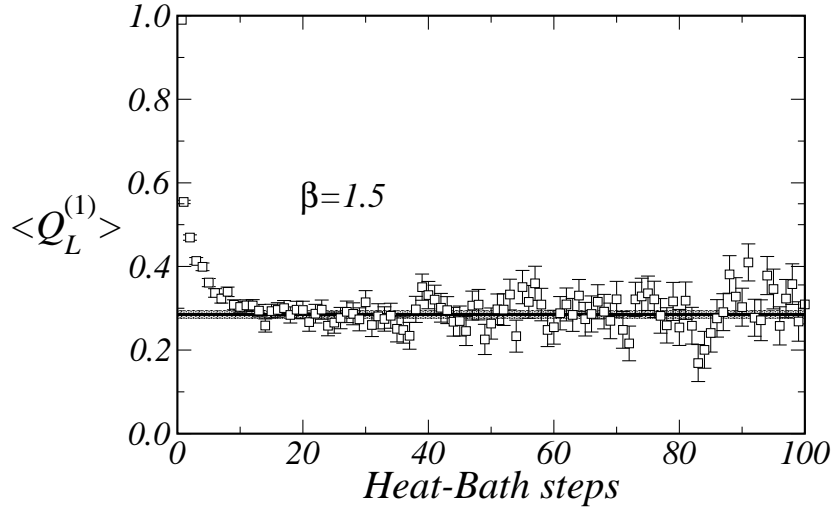}}
\caption{Data for $\langle Q_L^{(1)}\rangle$ start at $+1$
at the 0--th Heat--Bath step and
then they go down until reaching a plateau. The horizontal line and grey
band are the value and error respectively of $Z_Q^{(1)}(\beta=1.5)$.}
\label{Fig1}
\end{figure}

The determination of $Z_Q^{(1)}$ is not so trivial and an
example of such an evaluation is shown in Fig.\ref{Fig1}.
Measures of $Q_L^{(1)}$ on configurations that belong to
the topological sector $+1$ attain to a plateau (in general after a few
Heat--Bath steps) and stay on it for the rest of the updating steps.
The height of this plateau is the value of $Z_Q^{(1)}$.
In Table~1 the results for $Z_Q^{(1)}$ at the
values of $\beta$ used in the present work are given.

\vskip 1cm

The relevant consequence of the above considerations
for our study is that the vacuum angle $\theta$ is related to the
corresponding bare parameter by the expression $\theta=\theta_LZ_Q^{(1,2)}$
(which implies $\theta=\theta_L$ when $Q_L^{(2)}$ is used).

\section{Cluster algorithm for imaginary $\theta$}

Although the use of the topological charge density $Q_L^{(1)}$ requires
the knowledge of a renormalization constant, it brings about the advantage
that the action $S_L$ in~(\ref{latticeaction}) can be simulated on the lattice by
use of a fast cluster algorithm. Instead when the geometric charge $Q^{(2)}_L$
was used, the model was updated by a (rather slow) Metropolis algorithm.

Let us briefly describe the main characteristics of the new cluster
algorithm expressly devised for the present work.
The first part of an updating step with the usual Wolff algorithm~\cite{wolff}
for the standard O(3) sigma model without a $\theta$ term consists in choosing
a random unit vector $\vec{r}$ in such a way that every dynamical
field can be split in a component parallel to $\vec{r}$ and the rest,
$\vec{\phi}(x)=\left(\vec{\phi}(x)\cdot\vec{r}\right)\vec{r} +
\vec{\phi}_\bot(x)$, where $\vec{\phi}_\bot(x)$ denotes the part of
$\vec{\phi}(x)$ orthogonal to $\vec{r}$. Then the signs of
$\left(\vec{\phi}(x)\cdot\vec{r}\right)$ for all $x$ are updated \`a la
Swendsen--Wang as in the Ising model~\cite{swendsen}.

By introducing the above separation for $\vec{\phi}(x)$ in the
expression~(\ref{latticeQ1}) we can re--write it as
\begin{eqnarray}
&&Q_L^{(1)}(x)=\frac{1}{16\pi} \Big\{
  \left(\vec{\phi}(x)\cdot\vec{r}\right)
      \big(d_{1,2} + d_{-1,-2} + d_{2,-1} + d_{-2,1}\big) \nonumber \\
  &&+ \left(\vec{\phi}(x+\widehat{1}\,)\cdot\vec{r}\right)
      \left(d_{0,-2} - d_{0,2}\right)
   +\left(\vec{\phi}(x-\widehat{1}\,)\cdot\vec{r}\right)
      \left(d_{0,2} - d_{0,-2}\right) \nonumber \\
  &&+\left(\vec{\phi}(x+\widehat{2}\,)\cdot\vec{r}\right)
      \left(d_{0,1} - d_{0,-1}\right) 
   +\left(\vec{\phi}(x-\widehat{2}\,)\cdot\vec{r}\right)
      \left(d_{0,-1} - d_{0,1}\right) \Big\}\;,
\label{latticeQ1bis}
\end{eqnarray}
where $x\pm\widehat{1}$ means the site at the position one step forward
(backward) in the direction ``1'' starting from site $x$
and the notation $d_{i,j}$ stands for the $3\times 3$ determinant
(the three components for each vector must be unfold along the rows)
\begin{equation}
d_{i,j}\equiv \det\pmatrix{\vec{r}\cr
             \vec{\phi}(x+\widehat{i}\,\;)\cr
             \vec{\phi}(x+\widehat{j}\,\;)\cr}\; .
\label{definitiond}
\end{equation}
In this fashion the theory at each updating step looks like an Ising model
in the bosom of an external local magnetic field $h(x)$ because the
expression in Eq.(\ref{latticeQ1bis}) is linear in
$\left(\vec{\phi}\cdot\vec{r}\right)$. Recall that all Monte Carlo simulations
have been performed with an imaginary vacuum angle $\theta_L=+i\vartheta_L$,
($\vartheta_L\in{{\rm I}\kern-.23em{\rm R}}$).
By gathering all contributions
of the type shown in Eq.(\ref{latticeQ1bis}) that contain 
$\left(\vec{\phi}(x)\cdot\vec{r}\right)$ at site $x$
one can readily derive the effective magnetic field at this site,
\begin{eqnarray}
  h(x)&=&-\;\frac{\vartheta_L}{16\pi}\vert\vec{\phi}(x)\cdot\vec{r}\;\vert
\Big(d_{1,2} + d_{-1,-2} + d_{2,-1} + d_{-2,1} \nonumber \\
 &+& \;\;\, d_{-1,-1-2} + d_{-1+2,-1} + d_{1,1+2} + d_{1-2,1}
   \nonumber \\
 &+& \;\;\,d_{2,2-1} + d_{2+1,2} + d_{-2,-2+1} + d_{-2-1,-2}\Big)\;.
\label{fieldmag}
\end{eqnarray}
$d_{i+k,j}$ (and analogous terms in~(\ref{fieldmag}))
are the straightforward generalization of the above
definition~(\ref{definitiond}) when the site is obtained by shifting
two steps from the original position~$x$, the first in the direction
$\;\widehat{i}$ and the second in the direction $\widehat{k}\;$.

Hence the last step in the updating consists in applying to the
above expressions an algorithm valid for the Ising model in presence
of a magnetic field. In the literature there are two such algorithms,
the Lauwers--Rittenberg~\cite{lauwers} and the
Wang~\cite{wang,dimi} methods. After testing their
perfomances and comparing the corresponding decorrelation times
with the usual (multihit) Metropolis, Heat--Bath and
overHeat--Bath, we decided for the Wang algorithm. It consists in
placing the magnetic field on an extra, fictitious site
(called ghost site) that couples to every Ising spin through the value
of $h(x)$. Using this coupling on the same footing
as all other terms in the action, the Fortuin--Kasteleyn
clusters~\cite{fortuin} are
arranged by the Hoshen--Kopelman algorithm~\cite{hoshen} and
then updated with the usual $\frac{1}{2}$ probability.

Following the proof given in~\cite{wolff}, it can be seen that our
algorithm also satisfies the detailed balance property.

\section{Results}

Operators representing physical states can be built out
of an arbitrary
number of fundamental fields since supposedly the model is not parity invariant
for $\theta_L\not=0$. As the energy gap is given by the mass of a triplet
state~\cite{controzzi} we studied the correlation
functions of operators having one O(3) index as quantum number,
\begin{equation}
 {\overrightarrow{\cal O}}_1(x)\equiv\vec{\phi}(x)\,,\qquad
 {\overrightarrow{\cal O}}_2(x)\equiv\vec{\phi}(x)\times\vec{\phi}(x+\widehat{1}\,)\,.
\end{equation}
Then we calculated the related wall operators by averaging over the
$x_1$ coordinate (as usual $L$ is the lattice size),
${\overrightarrow{\cal W}}_i(x_2)\equiv \frac{1}{L}\sum_{x_1}
{\overrightarrow{\cal O}}_i(x)$ for $i=1,2$.

To single out the correct parity mixture for the physical particle and
to clean the signal from possible excited states, we extracted the triplet
mass $m$ by using the variational method of Ref.~\cite{kronfeld} where
the mass is obtained from the exponential decay of the
largest eigenvalue of the correlation matrix
$\langle {\overrightarrow{\cal W}}_i(x_2){\overrightarrow{\cal W}}_j(0)\rangle -
\langle{\overrightarrow{\cal W}}_i\rangle \langle{\overrightarrow{\cal W}}_j\rangle$.

\vskip 10mm
\begin{figure}[htbp]
\centerline{\epsfig{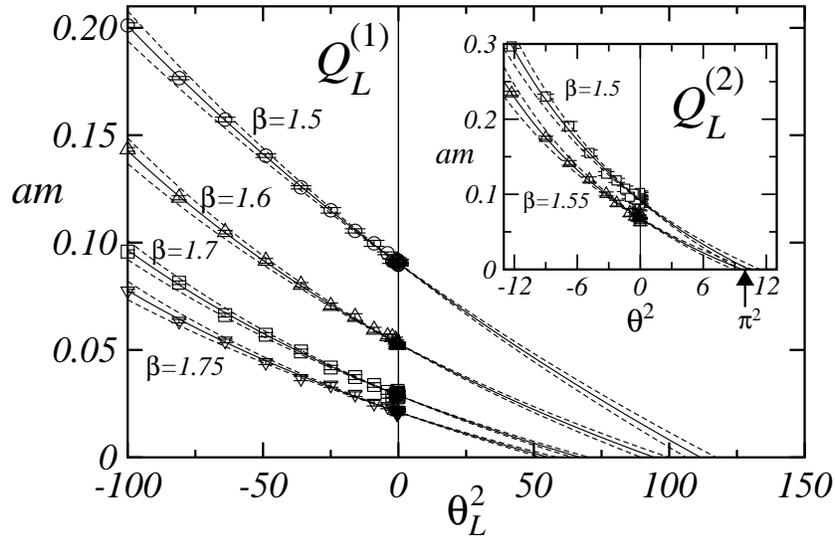}}
\caption{Behavior of the mass gap (in units of the lattice spacing $a$)
as a function of $\theta_L^2$. Main plot: circles ($\beta=1.5$),
up triangles ($\beta=1.6$), squares ($\beta=1.7$) and down triangles
($\beta=1.75$) are the data from the simulation at imaginary
$\theta_L$ ($\theta_L^2<0$) by using the $Q_L^{(1)}$ lattice
regularized topological charge. Each continuous line is the result of
the extrapolation described in the text and the dashed lines
enclose the boundary of its error. Inset: the same for the $Q_L^{(2)}$
regularization: squares ($\beta=1.5$) and up triangles ($\beta=1.55$).
In this case $\theta=\theta_L$ and the position of $\theta=\pi$ is indicated.}
\label{Fig2}
\end{figure}

\subsection{Results for $Q_L^{(1)}$}

$2\cdot 10^5$ decorrelated propagators were measured for all values of $\beta$ and
$\theta_L$. In the main plot of Fig.\ref{Fig2} the results for the triplet
mass are shown for four values of $\beta$.
The extrapolations in this figure were done by using
the functional form $(c_1 + c_2\, \theta_L^2)/(1 + c_3\, \theta_L^2)$.
We avoided using a functional form dictated by some theoretical argument,
like one that for real or imaginary $\theta_L$ goes like
$m(\theta_L)=c_1 (c_2^2 - \theta_L^2)^{2/3}$ which is, up to
logarithmic corrections, the Renormalization
Group prediction, because such an analytic form implicitly assumes
the vanishing of the mass at a precise value of $\theta$. Instead
we made the extrapolations with ratios of polynomials (which are
contemporaneously both simple and very general functional forms)
in order to leave room for any behaviour in the vacuum angle.
The results of the analytic continuations are given in Table~1.
The physical value of $\theta$ where the theory becomes gapless
is given by $\theta_{\rm end}=\theta_{L,{\rm end}} Z_Q^{(1)}$.
The numbers in the last column are in fair agreement
with the prediction that the model becomes massless when $\theta$ equals $\pi$.
Similar results (and $\chi^2$) were obtained from degree~2
or~3 polynomials in $\theta_L^2$, while ratios of higher order polynomials proved to be
statistically unlikely (their $\chi^2$ was too large).

\vskip 1cm

{{{\rm Table 1.} Values of $Z_Q^{(1)}$ and $\theta_{\rm end}$.}
\vskip 2mm
{\centerline{
\begin{tabular}{|p{10mm}|p{10mm}|p{14mm}|p{18mm}|p{16mm}|p{16mm}|} \hline
$\beta$ & $L$ & $\left(\theta_{L,{\rm end}}\right)^2$ & $Z_Q^{(1)}$ &
  $\chi^2/$d.o.f. & $\theta_{\rm end}$ \\ \hline
1.5 & 120 & 111(5) & 0.285(9) & 0.90 & 3.00(12) \\ \hline
1.6 & 180 & 94(5) & 0.325(6)  & 0.45 & 3.15(10) \\ \hline
1.7 & 340 & 67(3) & 0.380(6)  & 1.04 & 3.11(9)  \\ \hline
1.75 & 470 & 56(3) & 0.412(5) & 0.68 & 3.08(9)  \\ \hline
\end{tabular}
}}}

\vskip 1cm

The lattice sizes in Table~1 were chosen large enough to meet at $\theta_L=0$ the
condition $L/\xi\equiv L\cdot am\ge 10$. Once this inequality holds at
$\theta_L=0$, it is amply realized at the values of $\theta_L$ where
the simulations were performed as inferred from Fig.\ref{Fig2}. This fact
warrants the absence of significant finite size effects.

\subsection{Results for $Q_L^{(2)}$}

In this case a Metropolis
algorithm was used for updating and $10^5$ independent
propagators were measured for each value of $\theta$
(recall that in the present case $\theta_L=\theta$). We report
data for two values of $\beta$. They are displayed in the
inset of Fig.\ref{Fig2}. The value of $\theta$ where Haldane predicted
the closing of the mass gap is indicated with an arrow, $\theta^2=\pi^2$.
The numerical results are given in Table~2. Comments similar to
the $Q_L^{(1)}$ case apply to the extrapolations shown in the figure.
Again the results are in fair agreement with the conjecture.

\vskip 1cm

{{{\rm Table 2.} $\theta_{\rm end}$ for the operator $Q^{(2)}_L$.}
\vskip 2mm
{\centerline{
\begin{tabular}{|p{10mm}|p{10mm}|p{16mm}|p{16mm}|p{16mm}|} \hline
$\beta$ & $L$ & $\left(\theta_{\rm end}\right)^2$ &
  $\chi^2/$d.o.f.  & $\theta_{\rm end}$  \\ \hline
1.5 & 110 & 10.4(1.0) & 1.72  & 3.22(16) \\ \hline
1.55 & 150 & 9.7(1.0) & 0.73  & 3.11(16) \\ \hline
\end{tabular}
}}}

\vskip 1cm

By averaging all results for both topological charge operators and assuming
gaussian errors we obtain that the mass gap vanishes at $\theta_{\rm end}=3.10(5)$.

\section{Conclusions}

We have simulated the O(3) non--linear sigma model
in 2~dimensions with an imaginary $\theta$ term at several values of
the lattice coupling $\beta$. The mass gap was measured and extrapolated
towards real $\theta$. In all cases the extrapolation vanished at a
value of $\theta$ compatible with the Haldane conjecture $\theta=\pi$.
Our result is $\theta=3.10(5)$ which agrees within errors with the
conjecture. This value seems very robust as it is
independent of the topological charge density operator chosen for
the simulation. In particular, an operator $Q_L^{(1)}$ that requires
a non--trivial
renormalization constant leads to the same conclusion than another
operator (the geometric charge $Q_L^{(2)}$) that does not renormalize.

A new fast cluster algorithm was purposely introduced to simulate the
theory with an imaginary $\theta$ term. It works for the operator $Q_L^{(1)}$.
Instead, when the geometric topological charge $Q_L^{(2)}$ was used,
the theory was simulated by a (rather slow) Metropolis algorithm.

A salient outcome of our work is the good performance of the
analytic continuation from imaginary to real $\theta$. No theoretical
prejudices were assumed in the functional form used in the continuation,
apart from the obvious requirement that it is analytic. This can
be justified by a comparison with the phase diagram shown in
Ref.~\cite{bhanot}. In the case of the geometrical charge our largest
$\beta$ and $\theta=\theta_L$ (1.55 and 3.5 respectively) are very
far from the line of phase transitions; as for the $Q_L^{(1)}$ case,
our largest $\beta$ and $\theta=\theta_L Z_Q^{(1)}$ were 1.75 and
4.1, which again lie very far from any line of singular points.

The need to perform Monte Carlo simulations at imaginary
values of $\theta$ is actually a blessing in disguise since it forced
us to work at very small correlation lengths, as can be clearly
seen in Fig.\ref{Fig2}. Had we studied the theory directly at real
$\theta$ values, we would have met with severe finite size effects.

A key ingredient
for the successful extrapolation was to have got data from simulations
within a wide range of (imaginary) values of $\theta_L$ for all $\beta$,
($\vartheta_L\equiv-i\theta_L\in\left[0,10\right]$ when $Q_L^{(1)}$ was
used and $\vartheta_L=\vartheta\equiv -i\theta\in\left[0,3.5\right]$
for $Q_L^{(2)}$,
the difference of intervals being due to the effect of the non--trivial
renormalization that must be applied to
the former). All that looks encouraging for the numerical studies
based on the analytic continuation with respect to a parameter in
the theory, such as in QCD with non--zero chemical
potential~\cite{philipsen}.

\section{Acknowledgements}

We thank Ettore Vicari and Adriano Di Giacomo
for a critical reading of a preliminary draft of the paper and
Giuseppe Mussardo for illuminating conversations. We also want
to thank Gerrit Schierholz for valuable comments.
A.P. thanks Domenico Giuliano for useful discussions.


\end{document}